\documentclass[useAMS,usenatbib]{mnras} 
\usepackage{amsmath}
\usepackage{graphicx}
\usepackage[T1]{fontenc}
\usepackage{epstopdf}
\graphicspath{{../Figures/}}

\title[]{Local ultra faint dwarves as a product of Galactic processing during a Magellanic group infall}
\author[C. Yozin and K. Bekki]{C. Yozin\thanks{{\bf E-mail} 21101348@student.uwa.edu.au; kenji.bekki@uwa.edu.au} and K. Bekki\footnotemark[1]
\\\\
ICRAR, M468, The University of Western Australia, 
35 Stirling Highway, Crawley
Western Australia, 6009, Australia}

\begin{document}
\date{Accepted 2015 January. Received 2015 January; in original form 2015 January}
\pagerange{\pageref{firstpage}--\pageref{lastpage}} \pubyear{2013}
\maketitle
\label{firstpage}

\newcommand{\bia}{$b/a$}
\newcommand{\cnfw}{c$_{\rm NFW}$}
\newcommand{\ex}[1]{10$^{\rm #1}$}
\newcommand{\hi}{H~{\sc I}}
\newcommand{\ih}{$h^{\rm -1}$}
\newcommand{\fgas}{f$_{\rm gas}$}
\newcommand{\kms}{kms$^{\rm -1}$}
\newcommand{\lcdm}{$\Lambda$CDM}
\newcommand{\lsol}{L$^*$}
\newcommand{\maga}{mag arcsec$^{\rm -2}$}
\newcommand{\mstar}{M$_{\ast}$}
\newcommand{\mex}[2]{#1$\times$10$^{\rm #2}$}
\newcommand{\mhms}{M$_{\rm h}$/M$_{\ast}$}
\newcommand{\msol}{M$_{\odot}$}
\newcommand{\muc}{$\mu$(g,0)}
\newcommand{\mur}{$\mu$(g,r)}
\newcommand{\mvir}{M$_{\rm vir}$}
\newcommand{\pcsq}{pc$^{\rm -2}$}
\newcommand{\reff}{r$_{\rm e}$}
\newcommand{\rs}{r$_{\rm s}$}
\newcommand{\rvir}{r$_{\rm vir}$}
\newcommand{\rvirmw}{r$_{\rm vir,MW}$}
\newcommand{\vrot}{V$_{\rm rot}$}
\newcommand{\z}[1]{$z=#1$}

\begin{abstract}

The recent discoveries of ultra-faint dwarf (UFD) galaxies in the vicinity of the Magellanic system supports the expectation from cosmological models that such faint objects exist and are numerous. By developing a mass model of the Local Group and backwards integrating the Magellanic Clouds' present kinematics, we find that the locations of these UFDs are consistent with those predicted if previously associated with the Large MC as part of a loose association. We further demonstrate how these satellites are likely to have been processed by the Galactic hot halo upon accretion, with the implication that ongoing detections of extremely gas-rich objects on the periphery of the Galaxy and without clear stellar counterparts are analogous to the progenitors of the gas-deficient UFDs. Our model allows us predict the locations of other putative Magellanic satellites, and propose how their distribution/kinematics provide a novel constraint on the dynamical properties of the Galaxy. We also predict that the stripped metal-poor H~{\sc I}, previously associated with these UFDs, lies coincident with but distinguishable from the extensive Magellanic Stream.  

\end{abstract}


\section{Introduction}


The last decade has seen the proliferation of resolved or candidate satellites in the Local Group (LG), whether by wide-field imaging surveys and/or blind HI surveys \citep[e.g][]{simo07, mcco12, adam15}, in apparent fulfillment of the prediction for adundant dark substructure by the standard $\Lambda$CDM cosmological model \citep{klyp99}. Many constitute ultra faint dwarf (UFD) galaxies with, in some cases, as little as \ex{3-4}{}\msol{} in stellar mass and mass to light ratios of $>$1000 \msol/L$_{\odot}$. Ongoing and future observations of such systems provide great insight as to the baryonic processes dictating the faint end of the luminosity function \citep{brow15, meye15}.

These same observations can resolve another tension with the $\Lambda$CDM model, insofar as the spatial distribution of most LG satellites is not isotropic as expected but exhibits a correlation in phase-space \citep{krou05}; a notable recent interpretation of which is that these satellites are the tidal remnants of a past M31 merger event \citep{hamm13}.

An alternative explanation for this coherence lies in a group infall scenario, centred on the Large Magellanic Cloud \citep[LMC;][]{lynd76, bekk08, dong08}. Cosmological simulations of the local volume suggest Galactic-type halos preferentially accreted satellites from \ex{10-12}{} \msol{} subgroups 5 to 8 Gyr ago \citep{wetz15}, in which case the uncharacteristically blue and gas-rich Large MC is a prime candidate for a presumed subgroup central galaxy \citep{toll11}. Studies of known dwarf associations on the periphery of the LG \citep{tull06} suggest however an incompatibility between their distribution and that of the majority of LG satellites \citep{metz09}.

Recent independent detections of stellar overdensities in the DES survey by \citet[][K15]{kopo15}, \citet[][B15]{bech15} and \citet{kim15} have yielded the best candiates for Magellanic satellites thus far, with magnitudes $-6.6<{\rm M}_{\rm V}<-2.6$ and lying within a relatively small footprint close to the LMC. It is as yet uncertain if they fulfil an expectation from cosmological simulations that LMC-analog Galactic satellites would be accompanied by fainter objects lying in its vicinity of the sky \citep{sale11}. By identifying LMC-mass galaxies accreted onto MW-type halos in the ELVIS suite of N-body simulations, \citet{deas15} suggest that surviving LMC-type satellites account for as much as 7 percent of the LG inventory, although the specific association of these UFDs and others (e.g Leo II/IV/V) hinges on whether the LMC has just passed its first pericentre.

In this paper, we establish the likelihood that these UFDs were previously associated with the LMC as part of a loose association using the latest estimates for the dynamical/kinematic properties of the Galaxy/LMC, which affords the opportunity to consider group processing specific to the local environment. The absence of \hi{}-counterparts to these UFDs \citep{west15} leads to demonstrate, therefore, an evolutionary link between these stellar systems and the so-called gas-rich dark galaxy candidates, such as recently-discovered Leo P and AGC198606 located on the periphery of the Galactic halo \citep{adam15}.

\section{Orbital history of a hypothetical UFD progenitor}

\subsection{Method}

\begin{figure}
	\centering
	\includegraphics[width=1.\columnwidth]{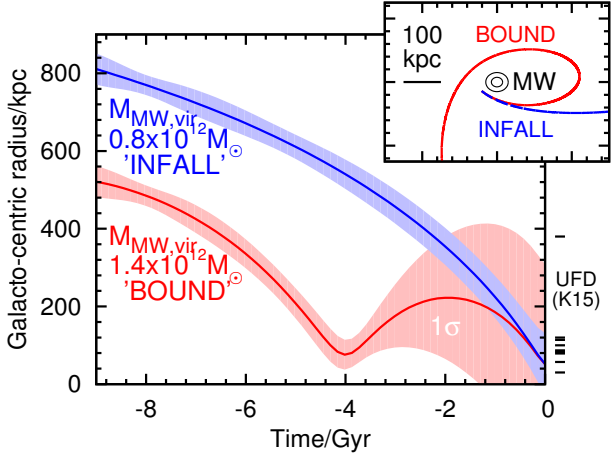}
	\caption{Galacto-centric radius of the LMC for orbits BOUND and INFALL, whose initial conditions differ only by the assumed present-day MW mass/$c$; the respective 1$\sigma$ spread in UFD model orbits with respect to the LMC are shown. The present radii of the observed UFDs (K15/B15) are shown with black dashes where the top-most corresponds to Eridanus 2; (Inset) LMC orbits in x-y space (coplanar with and centred on the MW disc).}
\end{figure}

We adopt a method that builds upon that of \citet{nich11}, in which a long-term MC-UFD association is deduced from tracing the trajectories of the LMC and small MC (SMC), with respect to the MW, back by 9 Gyr \citep[at which point infall is deemed to have commenced;][]{wetz15}, then integrating forwards \ex{5}{} UFDs models according to a Monte-Carlo technique, wherein their initial location is randomly sampled from an assumed Magellanic Group that is analogous to known dwarf associations \citep{tull06}. All orbits are integrated with the symplectic leap-frog algorithm. 

Using the HST third-epoch proper motions for the MCs \citep[][K13]{kall13}, we trace the past MC trajectories with a backwards-integration method composed of potential models of the LG \citep{mura80}, as adopted previously to establish the formation of the Magellanic Stream \citep[MS;][]{yozi14b}. Our model therefore consists of a MW comprising an adiabatically-contracted NFW dark halo with dynamical mass M$_{\rm MW}$, a Miyamoto-Nagai stellar disc with mass M$_{\rm d}$ and a Hernquist bulge with mass M$_{\rm b}$. The dynamical friction imposed on the orbits of the MCs while travelling through the MW halo is modelled according to the method established by \citet{zent03}. Our models of the MCs each consist of NFW halos with mass M$_{\rm LMC}$ (M$_{\rm SMC}$) and exponential discs with mass M$_{\rm d,LMC}$ (M$_{\rm d, SMC}$). 

The redshift evolution of the respective NFW mass and concentration ($c$) of the MW is incorporated here using merger rate-$z$ relations from \citet{fakh10} and \citet{muno11} and assuming the MW halo's virial radii (r$_{\rm vir}$) scales with mass M$_{\rm MW}$ as r$_{\rm vir}$$\propto$M$_{\rm vir}^{\rm 1/3}$. We also assume the stellar components of the MW, M$_{\rm d}$ and M$_{\rm b}$, have grown linearly by 50 percent since \z{2}{} \citep{snai14}. 

The stellar components of the MCs are fixed in mass, but their halo masses, M$_{\rm LMC}$ and M$_{\rm SMC}$, are assumed to lose mass as a function of time due to tidal truncation. The halo mass of each MC at 9 Gyr ago is thus prescribed the value that corresponds to their current stellar mass according to halo abundance matching methods \citep{behr13}; the mass loss rate that links these values with those deduced from present day observations (e.g. K13) is assumed constant.

Our model thus comprises various simplifications, many of which we find the results to be largely insensitive to. The key exception lies in the MW's total mass and $c$, insofar as they strongly influence the number of completed MW-MC orbits since infall. In this work, we consider a BOUND case corresponding to a high-mass MW (M$_{\rm MW}=1.6\times{\rm 10}^{\rm 12}$\msol{} \citep{gned10} and a INFALL case with a compact/low-mass MW (M$_{\rm MW}=0.8\times{\rm 10}^{\rm 12}$\msol{} \citep{kafl14}. While each of these estimates are based on stellar halo radial velocities, the respective assumptions for density/velocity anisotropy profiles vary significantly, leading to estimates of one and zero completed MW-MC orbits respectively.

In the absence of dynamical measurements of the observed UFDs, we infer their total masses from neighbouring LG objects in the size-luminosity plane \citep[e.g Segue 2/Leo V;][]{simo07, mcco12}, thereby adopting a stellar mass M$_{\rm d,UFD}$ of \ex{3.5}{}\msol{}. Abundance matching of subhalos in this faint regime is highly uncertain, relying on extrapolaton over 5 orders of magnitude; for our UFDs, modelled as NFW halos, we adopt a halo-to-stellar mass M$_{\rm h}$/M$_{\rm d}$|$_{\rm UFD}$ at 9 Gyr ago of \ex{5}{} (Fig. 2), as obtained from \citet{garr14} who fit their relations to the more luminous LG satellites.

The \z{0}{} properties of this LG model are summarised in Table 1. The thick lines in Fig. 1 illustrate the respective orbits of the LMC in both BOUND and INFALL scenarios as deduced from the backwards integration. The shaded regions that accompany them convey the 1$\sigma$ deviation in galactocentric radii ($r$) of the 10$^{\rm 5}$ UFDs as a function of time, whose respective orbits commence in the vicinity of the LMC at 9 Gyr ago. The initial UFD locations at this lookback time are randomly sampled assuming they trace the isotropic NFW distribution of DM in groups, with concentration $\sim$3 \citep{lin04} up to distances of $\sim$250 kpc from the central LMC \citep{tull06}, and with velocities constrained by an Eddington distribution function. 

\begin{figure}
\includegraphics[width=1.\columnwidth]{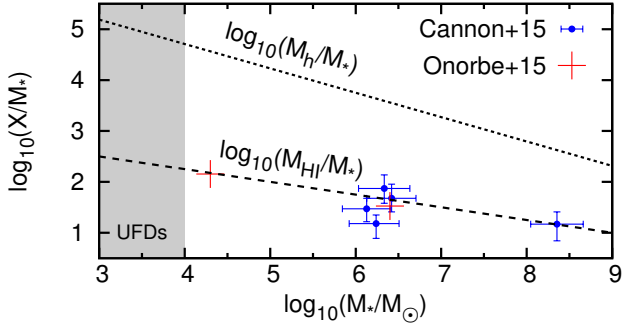}
\caption{Neutral hydrogen mass as a function of stellar mass for galaxies from the literature \citep{cann15, onor15}. Also shown is the halo-stellar mass relation from \citet{garr14}, derived from abundance matching and extrapolated to the region suspected to be occupied by UFD progenitors (grey shaded region).}
\end{figure}

\subsection{Results}

Fig. 1 illustrates how UFDs in a BOUND scenario are far more widely distributed in $r$ than in an INFALL scenario on account of significant scattering by Galactic tides upon first pericentre $\ge$4 Gyr. This result agrees qualitatively with that of \citet{deas15}; the significant proportion of UFD models that remain closely associated with the LMC in BOUND is also consistent with prior analytical predictions of their dispersion timescale \citep{bowd14}. 

At present, detections of such faint galaxies will be biased towards those at low $r$, at evidenced by the clustering of the K15/B15 sample at $r$$<$100 kpc. While it is therefore difficult to support a particular orbit scenario based on the observed UFD locations, Fig. 3 demonstrates our prediction for the location for further discoveries, where the modelled UFD locations at zero lookback time are compared with those lying in the DES footprint, in an equatorial projection. The probability distribution of UFDs as a function of $\delta$ reveals those in BOUND peak at low and high $r$, while those in INFALL are almost equally distributed along the (more than $\sim$100 kpc) span of the Magellanic Stream \citep{fox14}, many of which will likely be obscured by it.

\begin{figure}
	\centering
	\includegraphics[width=1.\columnwidth]{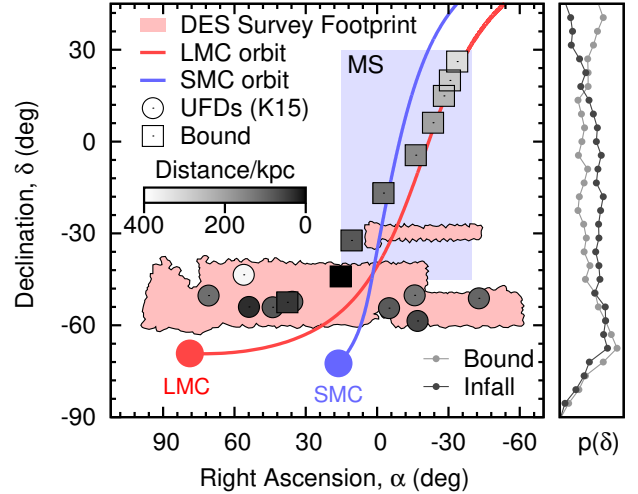}
	\caption{(Left panel) Present locations of the LMC/SMC (red/blue circles respectively) and their past orbits (BOUND scenario), projected in equatorial coordinates; also shown is the DES footprint (red shaded region), approximate location of the Magellanic Stream (MS; blue shaded region), and the present locations of observed UFDs (filled circles) coloured according to their heliocentric distance. Similarly, the average location of UFD models binned in distance increments of 40 kpc (up to 400 kpc) are shown with filled squares coloured according to their mean distance; (right) probability distribution of UFDs as a function of declination $\delta$}
\end{figure}

\begin{table} 
\centering
\caption{Summary of a) properties of observed UFDs; b) Initial UFD model parameters; c) present day MW and MC properties, where  M$_{\rm MW}$ and $c$ for the MW are given for BOUND and INFALL (in parentheses).}
\begin{tabular}{@{}lr@{}}
\hline
(a) UFD Parameters & Range \\
\hline
Half-light radius & 0.014 to 0.169 kpc \\
M$_{\rm V}$ & -6.6. to -2.7 mag \\
Heliocentric distance & 30 to 380 kpc \\
\hline
(b) UFD Model Initial Parameters & Value \\
\hline
Stellar mass M$_{\ast}$ & 10$^{\rm 3.5}$ \msol{} \\
Gas mass M$_{\rm g}$ & 10$^{\rm 6}$ \msol{} \\
DM halo mass M$_{\rm h}$ & 10$^{\rm 8.5}$ \msol{} \\
DM halo radius &  2.5 kpc \\
Stellar disc radius & 0.1 kpc \\
Gas disc radius & 0.26 kpc \\
\hline
(c) MW and MC Parameters & Value \\
\hline
MW halo mass M$_{\rm MW}$ & \mex{1.6(0.8)}{12}{} \msol{} \\
MW halo concentration, $c$ & 12 (21) \\
MW stellar disc mass & \mex{5.5}{10}{} \msol{} \\
MW stellar disc radius & 3.5 kpc \\
MW bulge mass & \mex{1}{10}{} \msol{} \\
MW bulge radius & 0.7 kpc \\
LMC (SMC) halo mass & \mex{5}{10}{} (\mex{3}{9}{}) \msol{} \\
LMC (SMC) halo concentration, $c$ & 9 (15) \\
LMC (SMC) virial radius & 120 (60) kpc \\
LMC (SMC) stellar disc radius & 1.4 (1.1) kpc \\
\hline
\end{tabular}
\end{table} 

\section{Gas loss to the hot galactic Corona}

Faint galaxies observed on the periphery of the MW halo and in extragalactic surveys (e.g. ALFALFA) report gas-to-stellar ratios of $>$100 \citep{huan12, cann15}, and the ongoing UV emission from others is proposed as a viable means for future discoveries \citep{meye15}. By contrast, the UFDs of K15/B15 (with the possible exception of the anomalously distant and blue Eridanus 2) appear to have no associated \hi{} \citep{west15}, and the strong possibility that they have been recently accreted to the Galaxy as part of a Magellanic Group would tentatively support their recent processing by this environment. 

However, the star formation histories of other Galactic satellites suggest that those with the least stellar mass were quenched at very high redshift. \citet{brow15} combine the photometry, spectroscopy and isochrones of six UFD Milky Way satellites to find in all cases an ancient stellar population, largely formed prior to \z{3}{}. \citet{weis15} define the epoch of quenching for each of their sample of 38 satellites as when 90 percent of their respective stars were formed, and establish a (possibly underestimated) quenched fraction of 0.4 by \z{2}{}, which in many cases predates their time of infall to the Galactic system. 

The pre-infall quenching is consistent with the extant gas of UFDs escaping their halos upon reionisation of the universe, but this scenario contrasts with ultra-high resolution simulations (sufficient to resolve M$_{\ast}\simeq{\rm 10}^{\rm 4}{}{\rm M}_{\odot}$ UFD analogues) that have recently probed the effect of the UV field on the UFD stellar mass regime and find it to merely suppress further gas accretion \citep{onor15}. The extended SF histories of several low mass dSphs highlight a further feasible scenario \citep{weis15}, wherein early quenching by the UV field is followed by gas accretion which renews a period of SF that is quenched once again upon infall. 

\subsection{Method}

In this section, we investigate the possibility that the absence of \hi{} associated with the K15/B15 UFDs can be explained in terms of an interaction between a previously gas-rich disc and the Galactic hot halo alone. We thus prescribe our UFD models with an initial \hi{}-to-stellar mass of \ex{2.5}{}; illustrated in Fig. 2, this is motivated by high M$_{\ast}$/L$_{\ast}$ systems imaged by the VLA \citep{cann15} and the simulated analogues from \citet{onor15}.

In an analytical method to establish the gas loss, we adopt the same 10$^{\rm 5}$ UFD orbits computed in Section 2, which in each case provides their respective galactocentric radius $r$ at a given lookback time. The MW's hot halo density at $r$ is computed from the modified-$\beta$ density model of \citet{mill15}; although the normalised hot halo mass distribution is fixed, its total mass is assumed to evolve in a self-similar manner that follows the evolving M$_{\rm MW}$ as documented in Section 2. 

The gas mass loss rate via ram pressure for each UFD orbit is predicated on the condition that:
\begin{equation}
	\rho{\rm V}^{\rm 2}\ge\Sigma({\rm r_{\rm d}})\frac{\delta\phi({\rm r}_{\rm d})}{\delta{\rm z}}|_{{\rm z}_0},
\end{equation}
where $\rho$ is the local halo density for a UFD with velocity V and a restoring force perpendicular to its disc computed from the gas surface density profile $\Sigma$ and potential gradient $\delta\phi/\delta{\rm z}$ at the in-plane radius r$_{\rm d}$. Thus we assume the stripping of all \hi{} beyond r$_{\rm d}$ for which the condition is met. 

A caveat with this method lies in the uncertainties regarding the height z$_{\rm 0}$ at which the gradient should be computed, and the rate at which stripped gas can fall back upon the galaxy. To address this issue quantitatively, we perform a single self-consistent N-Body/SPH ram pressure simulation for a single UFD orbit (in the BOUND scenario). By calculating the sum of the galaxy's SPH gas particles exceeding their respective escape speed within the galaxy's instantaneous potential, the corresponding gas stripping rate during the critical pericentre passages can be used to calibrate our analytical model. 

The details of this numerical apparatus is given in our previous studies \citep{bekk14, yozi15}, but can be briefly summarised as follows: a cubic lattice of SPH particles with periodic boundary conditions are used to model the Galactic hot halo local to the galaxy (up to 10 disc scalelengths). The self-gravitating galaxy model sits at rest within this lattice, while the density and relative velocity of the halo SPH particles are varied at each simulation timestep (length of order 10$^{\rm 5}$ yr) according to the aforementioned predefined UFD orbit and the $beta$-model density at the corresponding $r$.

Although the UFDs are not expected to be rotationally-supported, their analytical/simulation models adopt an exponential density model for an UFD progenitor with a size 0.1 kpc extrapolated from the tight size-mass relation (over $\rm 10^{\rm 7}<{\rm M}_{\ast}/{\rm M}_{\odot}<{\rm 10}^{\rm 11}$) from \citet{ichi12}, with a similar profile for the co-planar gas disc, albeit with a factor 2.6 larger scalelength \citep{krav13}.

The inclination of these discs with respect to its orbit is prescribed to an arbitrary $\sim$37.5 degrees. The analytical formula provided above is strictly defined for a face-on (0 degrees) inclination, although in simulations where the external ram pressure is comparable to the central pressure of the galaxy's gas disc (such as in the scenario considered here), \citet{roed06} find the ram pressure efficiency varies only for those inclinations close to edge-on (90 degrees). By assuming no preference in the infall inclination of the alleged Magellanic UFDs, we conservatively account for their simulated results by scaling our computed stripping rates by a factor 0.7. 

\subsection{Results}

The top row of Fig. 4 shows a snapshot from the simulation method taken during the first pericentre passage, illustrating the large \hi{} wake that trails the orbit of the UFD. The corresponding gas mass loss is shown in the second row, and compared with an equivalent UFD model from our analytical method and whose properties are calibrated by it; the peak \hi{} mass loss rate during the pericentre passages is $\sim$3-4$\times$10$^{\rm 5}$ M$_{\odot}$. The third row compares the gas loss associated with ram pressure for a single UFD model constrained to the LMC trajectory in both BOUND/INFALL scenarios. To account for up to magnitude variations in local halo densities around simulated stripped satellites \citep{bahe15}, we present for each orbit two runs in which the local $\rho$ is scaled by 1 and 10 respectively. 

This suite of models predict in general a (near) complete removal of extant \hi{} by \z{0}{} for those UFDs proximate in location to the LMC, consistent with \citet{west15}. The bottom row of Fig. 4 confirms this result for the \ex{5}{} UFD models in a log-scaled histogram of the gas-to-stellar ratio-heliocentric distance ($r$) phase space (where $\rho$ is scaled by random factor $f$, $0.1<f<10$). 

The large \hi{} loss rates reported above implies their remnants will be similar in both morphology and mass to the $\sim$\ex{5-6}{} \msol{} compact High Velocity Clouds that populate the MW halo \citep[e.g.][]{giov10}. For the instantaneous local density and velocity of the UFD, we can calculate the destruction timescale of a stripped \ex{5}{} \msol{} gas blob with effective size $d=0.1$ kpc and gas density $\rho_{\rm b}\sim{\rm 10}^{\rm 19}$cm$^{\rm -2}$ with the formula of \citet{blan07} i.e. $\tau_{\rm d}\simeq(\rho_{\rm b}/\rho)^{\rm 0.5}(d/V)$. 

If conservatively assuming a velocity $V$ equal to that of the UFD, we obtain $\tau$ of order several Gyr, implying that such blobs have feasibly survived to the present day since their removal up to 5 Gyr ago, independently of their progenitor UFD. While Fig. 3 shows how these blobs most likely lie in the vicinity of the MS, it is possible to distinguish them from fragments of the MS by their primordial metal abundances \citep[e.g][]{kuma15}, given that the chemical abdundance and filamentary structure of the Stream are consistent with a relatively metal-rich SMC as its origin \citep{fox13, yozi14b}.

The proximity between these the UFD orbits and the Stream also implies their mutual hydrodynamical interaction. It is beyond the scope of the methods presented thus far to examine this in detail, so we refer to the average gas density of the Stream (including a substantial ionised component), log($n_{\rm H}$cm$^{\rm -3}$)$\simeq$-1.8 \citep{fox14}, and a characteristic velocity dispersion of group associations \citep[$\sim$30 kms$^{\rm -1}$;][]{tull06} to represent the relative UFD-Stream velocity. The corresponding ram pressure ($\sim$$\rho$V$^{\rm 2}$) is approximately two orders of magnitude less than that imposed by the Galactic hot halo. Although seemingly insignificant, the suspected lifetime of the Stream deduced from tidal models \citep[e.g. 2 Gyr;][]{besl12} implies a long-term interaction that will necessitate an explicit modelling in more refined models.

\begin{figure}
\includegraphics[width=1.\columnwidth]{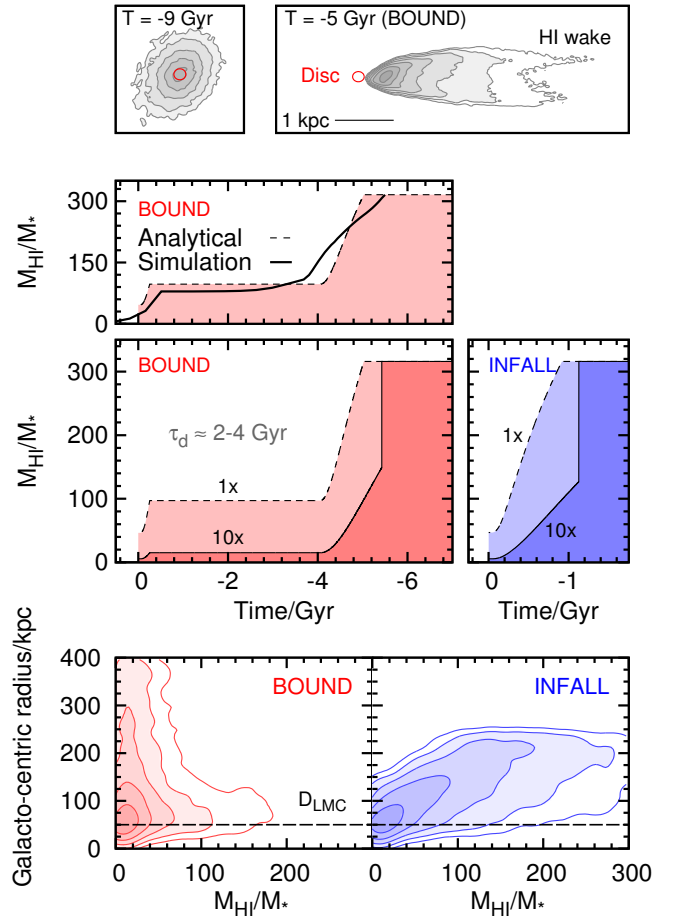}
\caption{(Top row) Selected snapshots from the N-Body/SPH simulation used to calibrate our analytical stripping model, where contours represent the gas surface density in log-scaling; (Second row) A comparison of the \hi{}-to-stellar mass as a function of time for a single UFD model in on a BOUND orbit as computed with an analytical method (dashed line), whose parameters are calibrated by the results of an equivalent N-Body/SPH simulation (solid line); (third row) The same, but for UFDs modelled with the analytical method only, and constrained to orbits BOUND and INFALL (left and right respectively), where hot halo density $\rho$ at a given $r$ is scaled by 1 and 10; (bottom row) The final galacto-centric radius of \ex{5}{} UFDs in both BOUND and INFALL scenarios, plotted as a log-scaled histogram (contours represent orders of magnitude) in the \hi-stellar mass/Heliocentric distance phase space.}
\end{figure}

\section{Discussion and conclusions} 


In this letter, we have presented a viable formation scenario for several of the ultra faint dwarf satellites recently detected by \citet{kopo15} and \citep{bech15}. Unlike the spatial distribution of dSphs at large in the LG, whose origin presently eludes clear explanation within the context of $\Lambda$CDM, these UFDs are consistent with their infall as part of a loose association centred on the LMC, a scenario that agrees well with those of cosmological simulations of the local volume \citep{wetz15}. 

If assuming the majority of these UFDs were originally Magellanic satellites, then our model yields a probability distribution for the present locations of these UFDs (Fig. 3) that predicts in the range of 10 more, most of which are lying within the Magellanic Stream. This represents the second prediction for stellar objects in this region, where previous attempts to detect the stellar arms postulated to have been tidally removed from the MCs \citep{yozi14b} have been thus far inconclusive \citep[e.g][]{osth97}. If such objects were found, they could be adopted as valuable distance constraints on the Stream itself.

It is postulated here that overcoming the detection biases that plague the current K15/B15 UFD sample, and revealing the wider population of putative Magellanic satellites, can provide constraints on their combined orbital history. By extension, this can provide a novel constraint on the wide range of Galactic dynamic masses reported in the literature; other such constraints have recently been provided by models of the tidal formation of the Sagittarius Stream \citep[yielding a surprisingly low mass estimate of 4$\times$10$^{\rm 11}$ M$_{\odot}$;][]{gibb14}. 

Our results generally compliment those of \citet{deas15}, who searched for surviving LMC-type satellites and their entourage within MW-type hosts of the ELVIS simulation. However, for the range of observationally-motivated MW masses considered in this study (${\rm 0.8-1.6}\times{\rm 10}^{\rm 12}$ \msol), the LMC is an unusual satellite for variety of reasons, not least its mass \citep[e.g.][]{rodr13}. For this reason, the scenario derived from their ELVIS sample is quantitatively different to ours, being biased towards ${\rm 2}\times{\rm 10}^{\rm 12}{\rm M}_{\odot}$ hosts. 

We note however that our non-static MW/MC model, like those of \citet{kall13}, is subject to a variety of simplifications that require further refinement. As noted by \citet{deas15}, the tidal processing of the UFDs prior to and during accretion remains to be considered, as emphasised not only by the morphological disruption of other UFDs \citep{simo07}, but also the potentially long-term LMC-SMC binarity as part of a common halo \citep{bekk08, besl12}. Incidentally, we note that although ram pressure acts directly on the UFDs' gas, our single N-Body/SPH simulation demonstrated how the significant shift in baryons for such large M$_{\rm HI}$/M$_{\ast}$ ($\sim$100) drives a morphological change manifesting in the elongation of the disc in the direction of its \hi-wake.

We have assumed the effects of reionisation in the UFD mass regime are separable to those as part of group processing \citep{wetz15}, and thus do not preclude large gas masses upon infall \citep{onor15}. Therefore, the disparity between dark gas-rich galaxies \citep[without clear stellar counterparts at present, e.g.][]{toll15}, and the paucity of gas detected near the stellar overdensities identified by K15 and B15 \citep{west15}, can be traced to the highy eccentric/energetic trajetory of the Magellanic Clouds/Group which penetrates some of the densest regions of the Galactic hot halo. 

By accounting for the possible influence of the Magellanic Stream, the primary uncertainty pertaining to our ram pressure model lies in the gas stripping rate. Nonetheless, our second key prediction lies in observable signatures of this mechanism, in the form of compact HVCs lying in the wake of the Magellanic Stream and distinguishable by their low metallicity. Accordingly, the UFDs objects are expected to be dominated by an old stellar population with a negligible specific star formation since commencement of their recent accretion between 1 and 5 Gyr ago. 

\section*{Acknowledgements}

We thank the referee, Joss Bland-Hawthorn, for his comments which improved this manuscript. CY is supported by the Australian Postgraduate Award Scholarship. 

\bibliographystyle{mn2e}
\bibliography{bib}

\bsp
\label{lastpage}
\end{document}